%% file: main.tex
\documentclass[conference, 10pt, a4paper]{IEEEtran}
\IEEEoverridecommandlockouts
\input{macro}

\urldef\myurl\url{{*qi.liao, $^\dagger$qiangliu}@nokia-bell-labs.com}

\begin{document}

\title{Inter-Cell Slicing Resource Partitioning via Coordinated Multi-Agent Deep Reinforcement Learning}

\author{
	\IEEEauthorblockN{Tianlun Hu\IEEEauthorrefmark{1}\IEEEauthorrefmark{4}, Qi Liao\IEEEauthorrefmark{1}, Qiang Liu\IEEEauthorrefmark{2}, Dan Wellington\IEEEauthorrefmark{3}, and Georg Carle\IEEEauthorrefmark{4}}
	\IEEEauthorblockA{\IEEEauthorrefmark{1}Nokia Bell Labs, Stuttgart, Germany\\
	\IEEEauthorrefmark{2}School of Computing, University of Nebraska-Lincoln,  United States\\
	\IEEEauthorrefmark{3}Nokia Software, Bellevue, United States\\
	\IEEEauthorrefmark{4}Dept. of Informatics, Technical University of Munich, Germany\\
	\{\IEEEauthorrefmark{1}\IEEEauthorrefmark{4}tianlun.hu,\IEEEauthorrefmark{3}dan.wellington\}\url{@nokia.com}, \IEEEauthorrefmark{1}qi.liao\url{@nokia-bell-labs.com,}\IEEEauthorrefmark{2}qiang.liu\url{@unl.edu,}\IEEEauthorrefmark{4}\url{carle@net.in.tum.de}}
}

\maketitle

\begin{abstract}
	Network slicing enables the operator to configure virtual network instances for diverse services with specific requirements. To achieve the slice-aware radio resource scheduling, dynamic slicing resource partitioning is needed to orchestrate multi-cell slice resources and mitigate inter-cell interference. It is, however, challenging to derive the analytical solutions due to the complex inter-cell interdependencies, inter-slice resource constraints, and service-specific requirements. In this paper, we propose a multi-agent deep reinforcement learning (DRL) approach that improves the max-min slice performance while maintaining the constraints of resource capacity. We design two coordination schemes to allow distributed agents to coordinate and mitigate inter-cell interference. The proposed approach is extensively evaluated in a system-level simulator. The numerical results show that the proposed approach with inter-agent coordination outperforms the centralized approach in terms of delay and convergence. The proposed approach improves more than two-fold increase in resource efficiency as compared to the baseline approach.
\end{abstract}

\vspace{-.5ex}
\makeatletter{\renewcommand*{\@makefnmark}{}\footnotetext{This work was supported by the German Federal Ministry of Education and Research (BMBF) project KICK [16KIS1102K].}\makeatother}

\section{Introduction}
\begin{figure}[t]
	\centering
	\includegraphics[width=0.45\textwidth]{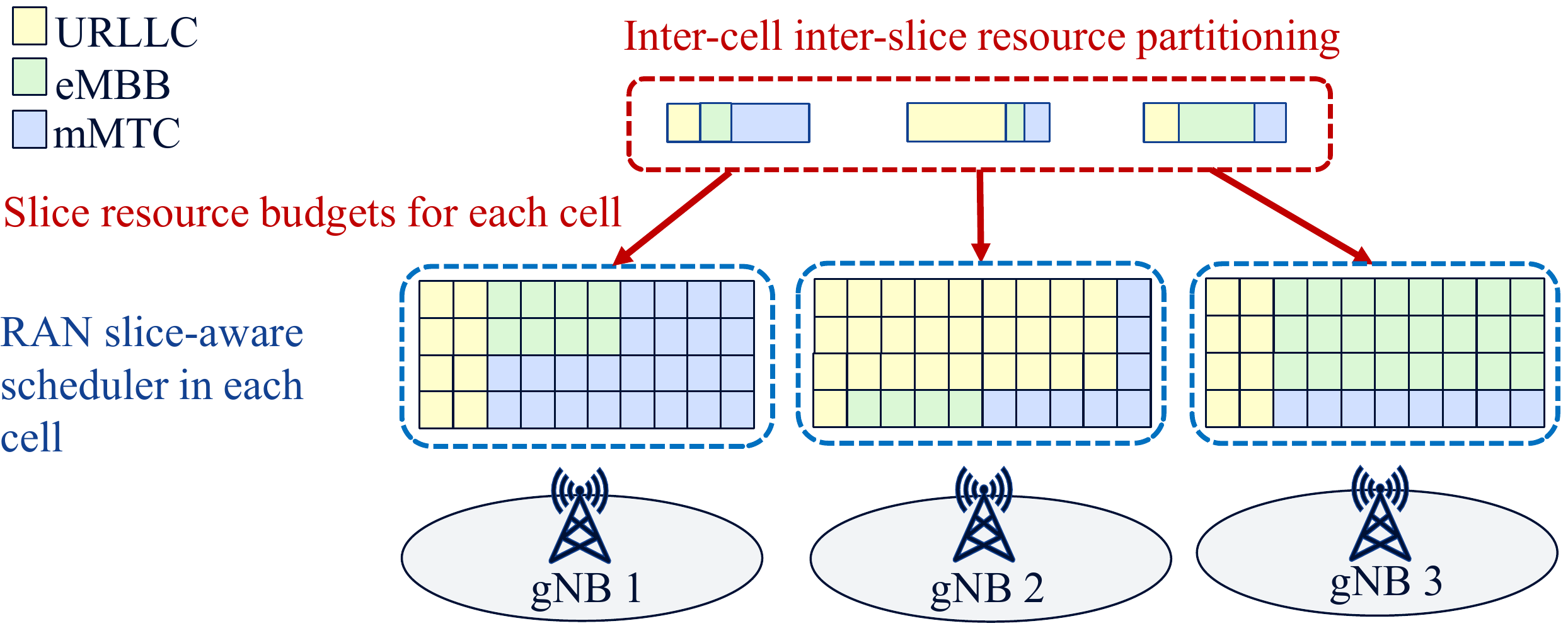} 
	\caption{Dynamic slicing inter-cell inter-slice resource partitioning}
	\label{fig:RANSlicing}
\end{figure}

Network slicing enables the network operator to create isolated virtual networks (\emph{aka.} slices) based on the common network physical infrastructures.
The network slices can be customized to support diverse use cases and services, e.g., enhanced mobile broadband and ultra reliable low latency communications, with heterogeneous performance requirements such as throughput and latency.
To satisfy the performance and coverage requirements of slices, the network operator aims to partition the radio resources, e.g., physical resource blocks (PRBs), in multiple base stations such as gNBs, as shown in Fig. \ref{fig:RANSlicing}.
The objective is to meet the performance requirements of distinct slices with minimal inter-cell resource usage and thus maximal resource efficiency.

Existing model-based solutions formulate the resource partition problem with mathematical models and solve the problem with various optimization techniques, e.g., linear programming\cite{Addad2020OptimizationMF}, \cite{Beshley2021QoSAwareOR} and convex optimization\cite{Fossati2020MultiResourceAF}, \cite{Ma2020SlicingRA}.
For example, Addad \emph{et. al.} \cite{Addad2020OptimizationMF} formulated the network function deployment problem as \ac{MILP} under the constraints of resource, latency and bandwidth, and proposed a heuristic algorithm to solve the problem.
Cavalcante \emph{et. al.} \cite{8675967} formulated a max-min fairness problem to control the load-coupled interference in wireless networks, and the problem is transformed into a fixed point problem that can be efficiently solved by existing low complexity iterative algorithms.  
These solutions fail to achieve the optima in real networks because the approximated models cannot fully represent the complex networks. 

Recently, the model-based solutions, especially deep reinforcement learning (DRL), show a very promising potential on automatically learn to manage radio access networks without the need of prior models.
For example, Liu \emph{et. al.} \cite{ConstrainedRLNetSlicing} proposed an adaptive constrained reinforcement learning algorithm based on interior-point policy optimization (IPO) in the scenario of a single base station. 
Liu \emph{et. al.} \cite{DeepSlicingQiang} designed a DeepSlicing algorithm to allocate the resource to different slices, where each slice is associated with a DRL agent and a coordinator is created to coordinate the resource capacity in the base station.
However, these works are designed to address the resource allocation problem in single cell scenario.
In \cite{Alqerm2016ACO} and \cite{Zhao2019DeepRL}, the authors proposed \ac{DRL} solutions with discrete action space for multi-cell scenarios but the achievable performance is limited due to the discrete resource partitioning actions. The authors in \cite{Song2021ADR} and \cite{Peng2020DeepRL} introduced resource management system with continuous \ac{DRL} for complex scenarios, however, none of them addressed the inter-cell interdependencies and inter-slice resource constraints.
As the network deployment becomes denser, which causes more severe inter-cell interference among a large number of cells, there is a need for coordinated multi-agent \ac{DRL} design capable of capturing complex inter-cell and inter-slice interactions with low model complexity. 

In this paper, we investigate the resource partition problem in network slicing under the multi-cell scenario.
We aim to improve the max-min slice performance while satisfying the constraints of resource capacity.
To tackle the inter-cell interference, we propose a multi-agent DRL approach including two coordination schemes, i.e., with or without inter-agent coordination.
Moreover, we develop two methods to handle the constraints of instantaneous resource capacity in each agent. 
The contributions of this paper are summarized as follows:
\begin{itemize}
    \item We formulate the dynamic inter-cell slicing resource partitioning problem to improve the max-min slice performance while meeting the constraints of resource capacity. 
    \item We propose a multi-agent \ac{DRL} approach to solve the problem with two coordination schemes, i.e., with and without {\it inter-agent coordination}. We show that inter-agent load sharing improves the performance of the distributed scheme, while allowing the lower model complexity and faster convergence compared to the centralized \ac{DRL} approach. 
    \item We develop two methods, i.e., {\it reward shaping} and {\it decoupled softmax embedding}, to let the \ac{DRL} agent aware of the resource constraints.
    \item We evaluate the proposed solutions with a system-level simulator and show that the inter-agent coordination scheme outperforms the centralized approach in terms of slice performance, while achieving more than two-fold increase in resource efficiency compared to the traffic-aware baseline.
\end{itemize}

This paper is organized as follows. We define the system model in Section \ref{sec:model} and formulate the inter-cell inter-slice resource partitioning problem in Section \ref{sec:problem}. In Section \ref{sec:Approaches} we propose the distributed \ac{DRL} solutions to the problem including two schemes with and without coordination. The numerical results are provided in Section \ref{sec:numerical}. Finally, we conclude this paper in Section \ref{sec:concl}.

\section{System Model}\label{sec:model}
We consider a network system consisting of a set of cells $\K:=\left\{1, \ldots, K\right\}$ and a set of slices $\N:=\left\{1,  \ldots, N\right\}$. Each slice $n$ has pre-defined throughput requirement $\phi_n^\ast$ and delay requirement $d_n^\ast$. The system runs on discrete time slots $t\in\NN_0$.  To adapt to the time varying network traffic and satisfy the slice-aware service requirements in terms of both throughput and delay, the \ac{OAM} adjusts the inter-slice resource partitioning for all cells periodically. The optimized slicing resource partitions are provided to the \ac{RAN} scheduler in each cell, and used by the scheduler as the slicing resource budget for the further \ac{PRB} allocation at a finer time-granularity (as shown in Fig. \ref{fig:RANSlicing}).  

Considering the temporal and inter-cell interdependencies, we model the multi-cell system as a \ac{MDP} defined by the tuple $\left(\Ss, \A, P(\cdot), r(\cdot), \gamma\right)$, where $P:\Ss\times \A \times \Ss \to [0, 1]$ indicates the transition dynamics by a conditional distribution over the state space $\Ss$ and the action space $\A$, $r:\Ss\times \A\to\R$ denotes the reward function, and $\gamma\in[0,1]$ is the discount factor. 

The {\bf state} at time slot $t$, denoted by $\vs(t):=[\vs_1(t), \ldots, \vs_K(t)]\in\Ss$, is an observation of the entire system, where $\vs_k(t)\in\Ss_k$ is the local state observed from cell $k$.
The {\bf action} at slot $t$, denoted by $\va(t):=[\va_1(t), \ldots, \va_K(t)]\in\A$, includes the resource partitioning to each slice and each cell $a_{k,n}(t) \in [0, 1]$, for $k\in\K, n\in\N$. We further introduce a \lq\lq headroom\rq\rq \ (or reserved bandwidth) to the allocated resource for two reasons: 1) improve the resource efficiency, and 2) to convert the inequality action constraints to the equality ones. Let the headroom in cell $k$ be denoted by $a_{k,0}(t)\in[0, 1]$. The local action is then defined as  $\va_k(t) := [a_{k,0}(t),  \ldots, a_{k,N}(t)]\in\A_k$. Given the inter-slice resource constraints in each cell, the local action space $\A_k$ and the global action space $\A$ yield
\begin{align}
\A_k & := \left\{\va_k\bigg|a_{k,n}\in[0,1], \forall n\in\N\cup\{0\}; \sum_{n=0}^{N} a_{k,n} = 1\right\} \label{eqn:local_actionspace}\\
	\A & := \left\{\va\big|\va_k\in\A_k, \forall k\in \K\right\}. \label{eqn:global_actionspace}
\end{align}

Our objective is to satisfy the throughput and delay requirements$(\phi_n^\ast , d_n^\ast)$ for every slice $n\in\N$ and every cell $k\in\K$. Thus, given the observed average throughput $\phi_{k,n}(t)$ and average delay $d_{k,n}(t)$ at slot $t$ for each slice $n$ and cell $k$, we define the {\bf reward} function as below:
\begin{equation}
	r(t) := \min_{k\in\K, n\in\N} \min\left\{\frac{\phi_{k,n}(t)}{ \phi_{n}^\ast}, \frac{d_n^\ast}{ d_{k,n}(t)}, 1\right\}.
	\label{reward}
\end{equation}
Reward \eqref{reward} means that if any per-slice throughput or delay in any cell does not meet the requirement, we have $r(t)<1$. Otherwise, if all requirements are met, the reward is upper bounded by $1$. Note that the second term $d_n^\ast/d_{k,n}(t)$ is {\bf inversely} proportional to the actual delay, namely, if the delay is longer than required, this term is smaller than $1$. 

\section{Problem Formulation}\label{sec:problem}
Our problem is to find the policy $\pi:\Ss\to\A$, which decides the inter-cell inter-slice resource partitioning $\va\in\A$ based on the observation of network state $\vs\in\Ss$, to maximize the expectation of the cumulative discounted reward defined in \eqref{reward} of a trajectory for a finite time horizon $T$. The problem is given by:
\begin{problem}
\label{prob:RL}
\begin{equation}
  \max_{\pi} \  \Ex_{\pi} \left[\sum_{t=0}^{T} \gamma^t r\big(\vs(t), \va(t) \big)\right] 
  \mbox{s.t. } \va\in\A,
  \label{eqn:problem}
\end{equation}
where $\A$ is defined by \eqref{eqn:local_actionspace} and \eqref{eqn:global_actionspace}.
\end{problem}

The challenge of solving the above-defined problem are two-fold. Firstly, the reward function \eqref{reward} depends on high-dimensional global state and action spaces and involves complex inter-agent dependencies. For example, increasing resource partition in one slice $n$ and cell $k$ improves its own service performance, however, it decreases the available resource allocated to other slices in the same cell and increases the interference received in the neighboring cells, which may further result in a general service degradation. The second challenge is caused by the intra-cell inter-slice resource constraints \eqref{eqn:local_actionspace}. Although various methods are proposed to solve the constrained \ac{MDP} problems, e.g., by using Lagrangian method \cite{paternain2019constrained} or Projection-based Safety layer \cite{dalal2018safe}, there still exists the problem of oscillations and overshooting caused by constraint-violating behavior during agent training.

\section{Proposed Approaches}\label{sec:Approaches}
%
%
In this section, we first present the distributed multi-agent \ac{DRL} approach in terms of two different schemes to solve Problem \ref{prob:RL}: {\it distributed scheme without coordination}, and {\it distributed scheme with inter-agent coordination}. Then, we briefly introduce the actor-critic method to solve the \ac{DRL} problem. Last but not least, we propose two methods to deal with the inter-slice resource constraints.
%
\subsection{Proposed Distributed Schemes} \label{subsec:schemes}

\subsubsection{Distributed Multi-Agent Scheme without Coordination}
\label{sssec:Distributed}
The distributed approach allows each agent to learn a possibly different model and make its own decision on the local action, based on local or partial observation. In contrast to the conventional centralized approach, the distributed approach may not achieve the performance as good as the centralized one due to the limited observation. However, it may converge much faster and be more sample efficient by using a less complex model based on local states and actions.

We first consider the distributed approach without coordination, i.e., each agent $k$ only observes its {\bf local state} $\vs_k$. In particular, we include the following measurements and performance metrics  into the state $\vs_k$ for each cell $k\in\K$:
\begin{itemize}
  \item Average per-slice user throughput $\left\{\phi_{k,n}: n\in\N\right\}$;
  \item Per-slice load $\left\{l_{k,n}: n\in\N\right\}$;
  \item Per-slice number of active users $\left\{u_{k,n}: n\in\N\right\}$.
\end{itemize}
Thus, with the above defined three slice-specific features, the local state $\vs_k$ has a dimension of $3N$.

Each agent $k$ computes a {\bf local reward} $r_k$, and makes decision on the {\bf local action} $\va_k\in\A_k\subset [0, 1]^{N+1}$. The local reward based on the local observations is computed by 
\begin{equation}
	\label{reward_dis}
	r_k(t) := \min_{n\in\N} \min\left\{\frac{\phi_{k, n}(t)}{ \phi_{n}^\ast}, \frac{d_n^\ast}{d_{k,n}(t)}, 1\right\}.
\end{equation}

Each agent trains an independent model without communicating to others. Note that $r_k$ depends not only on the local state-action pair, but also on the states and actions of other agents, and we have $r(\va(t), \vs(t)) = \min_{k\in\K}r_k(\va(t), \vs(t))$. Thus, the distributed scheme approximates $r_k(\vs, \va)$ with $\tilde{r}_k(\vs_k, \va_k)$, decomposes Problem \ref{prob:RL} with $K$ independent subproblems, and finds the following local policies $\pi_k:\Ss_k\to\A_k$, $\forall k\in\K$:
\begin{equation}
    \label{eqn:local_policy}
    \pi_k^{\ast} = \argmax_{\pi_k; \va_k\in\A_k} \Ex_{\pi_k} \left[\sum_{t=0}^{T} \gamma^t \tilde{r}_k\big(\vs_k(t), \va_k(t) \big)\right], \forall k\in\K.
\end{equation}

The disadvantage of \eqref{eqn:local_policy} is that, because $r_k(\va(t), \vs(t))$ are strongly coupled to the joint actions and states of all neighboring agents, the approximation $\tilde{r}_k\big(\vs_k(t), \va_k(t) \big)$ based on the local observations can be erroneous, which may result in poor learning performance. 
%
%
\subsubsection{Distributed Multi-Agent Scheme with Inter-Agent  Coordination}
\label{sssec:Distributed_Comm}

%
In recent years, a promising direction of {\it distributed learning with inter-agent coordination} has attracted much attention \cite{foerster2016learning}. Allowing the agents to communicate for acquiring a better estimate of the global state improves the performance of the distributed method, while remaining the low complexity of the learning model. 

To help the distributed agents better estimate $r_k(\va(t), \vs(t))$ and capture the inter-agent dependencies, we propose to let the agents communicate and exchange additional information. Let each agent $k$ sends a message $\vm_k$ to a set of its neighboring agents, denoted by $\K_k$. Then, each agent $k$ holds the following information: local state and action pair $(\vs_k, \va_k)$ and received messages $\overline{\vm}_k:=\left[\vm_i: i\in\K_k\right]$.

One option is to directly use all received messages $\overline{\vm}_k$ along with $(\vs_k, \va_k)$ to estimate $r_k(\vs, \va)$ with $\tilde{r}_k(\vs_k, \overline{\vm}_k, \va_k)$. However, if the dimension of the exchanged message is high, this increases the complexity of the local model.

An alternative is to extract from the received messages $\overline{\vm}_k\in\R^{Z^{(m)}}$ useful information $\vc_k\in\R^{Z^{(c)}}$ with $g:\R^{Z^{(m)}}\to\R^{Z^{(c)}}:\overline{\vm}_k\mapsto\vc_k$, such that $Z^{(c)} \ll Z^{(m)}$, where $Z^{(m)}$ and $Z^{(c)}$ stand for the corresponding dimension. We can then use $\tilde{r}_k(\vs_k, \vc_k, \va_k)$ to approximate $r_k$, by capturing the hidden information in the global state, while remaining low model complexity. Pioneer works such as \cite{foerster2016learning} proposed to learn the extraction of the communication messages by jointly optimizing the communication action with the reinforcement learning model. However, for practical systems, the jointly training of multiple interacting models can easily result in unstable convergence problems. To provide a robust and efficient practical solution, we want to leverage the expert knowledge to extract the information. Knowing that the inter-agent dependencies are mainly caused by the load-coupling inter-cell interference, we propose to let each agent $k$ communicate with its neighboring agent the slice-specific load information $l_{k,n}$, $\forall n\in\N$. Then, based on the exchanged load information, we simply compute the average per-slice neighboring load as the extracted information $\vc_k(t)$. Namely, we define a deterministic function
\begin{equation}
  \label{neighbor_info}
  \begin{aligned}
  g_k: & \R^{N|\K_k|}\to\R^N : [l_{i, n}:n\in\N, i\in\K_k] \mapsto \vc_k(t)\\
    \mbox{with } &  \vc_k(t):=\left[\frac{1}{|\K_k|}\sum_{i\in \K_k}l_{i,n}(t): n\in\N \right].
   \end{aligned}
   \vspace{-1ex}
\end{equation}

Therefore, the proposed scheme is to find the following local policies $\pi_k:\Ss_k\times\R^N\to\A_k$ with distributed \ac{DRL} agents $k\in\K$:
\begin{equation}
\vspace{-1ex}
    \label{eqn:local_comm_policy}
    \pi_k^{\ast} = \argmax_{\pi_k; \va_k\in\A_k} \Ex_{\pi_k} \left[\sum_{t=0}^{T} \gamma^t \tilde{r}_k\big(\vs_k(t),\vc_k(t), \va_k(t) \big)\right], \forall k\in\K.
\vspace{-1ex}
\end{equation}

\subsection{Actor-Critic Method}\label{subsec:ActorCritic}
%
%
We consider to solve the \ac{DRL} problems with actor-critic approaches \cite{Konda1999ActorCriticA}, because of its effectiveness when dealing with high dimensional and continuous state and action spaces. Such approaches solve the optimization problem by using critic function $Q(\vs_t, \va_t|\theta^Q)$ (in this subsection, we denote $\vs(t)$ and $\va(t)$ by $\vs_t$ and $\va_t$ respectively for brevity) to approximate the value function, i.e., $Q(\vs_t, \va_t|\theta^Q) \approx Q^\pi(\vs_t, \va_t)$, and actor $\pi(\vs_t|\theta^\pi)$ to update the policy $\pi$ at every \ac{DRL} step in the direction suggested by critic.


In this work, we use \ac{TD3} \cite{Fujimoto2018AddressingFA} as off-policy \ac{DRL} algorithm built on top of the actor-critic methods. As the extension of \ac{DDPG} \cite{Silver2014DeterministicPG}, \ac{TD3} overcomes the \ac{DDPG}'s problem of overestimating Q-values by introducing twin critic networks for both networks $Q_{\theta_1}, Q_{\theta_2}$ and target networks $Q_{\theta'_1}, Q_{\theta'_2}$. The actor is updated by policy gradient based on the expected cumulative reward $J$ with respect to the actor parameter $\theta^\pi$, as:
\begin{equation}
\begin{aligned}
	\nabla_{\theta^\pi}J &\approx \Ex\left[\nabla_{\theta^\pi}Q(\vs, \va|\theta^Q)|_{\vs=\vs_t, \va=\pi(\vs_t|\theta^\pi)}\right]\\
	&= \Ex\left[\nabla_{\va}Q(\vs, \va|\theta^Q)|_{\vs=\vs_t, \va=\pi(\vs_t)}\nabla_{\theta^\pi}\pi(\vs_t|\theta^\pi)\right].
\end{aligned}
\end{equation}
The critic parameter $\theta^Q$ is updated with temporal difference learning, given by:
\begin{equation}
\begin{aligned}
	L\left(\theta^Q\right) & = \Ex\left[\left(g_t - Q(\vs_t, \va_t|\theta^Q)\right)^2\right], \\
\mbox{where }	g_t & = r_t + \gamma Q\left(\vs_{t+1}, \pi\left(\vs_{t+1}|\theta^{\pi}\right)|\theta^{Q}\right).
	\end{aligned}
\end{equation}

\subsection{Methods to Deal with Resource Constrains}\label{subsec:constrant}
We compare two solutions to address the inter-slice resource constraints in \eqref{eqn:local_actionspace}: the first is to reshape the reward function with additional term to penalize the violation of the resource constraints, and the second is to reconstruct the network architecture with additional regularization layer. 

\subsubsection{Reshaping the Reward Function}
\label{sssec:Penalty}
We add a penalty term to the original reward function \eqref{reward} to penalize the actions violating the constraints $\sum_{n=0}^{N} a_{k,n} = 1, \forall k \in\K$. At time slot $t$, the penalty is defined by the allocated resource ratio exceeding the maximum quota. The modified reward function with penalty term is defined as:
\begin{equation}
\begin{aligned}
	\label{reshaping}
	r(t) :=  \min_{k\in\K, n\in\N} &  \min\Bigg\{\frac{\phi_{k,n}(t)}{ \phi_{n}^\ast}, 
	\frac{d_n^\ast}{ d_{k,n}(t)}, 1\Bigg\} - \beta h_{k,n}(t) \\
	\mbox{with } h_{k,n}(t) := & \left|1 - \sum_{n=1}^{N} a_{k,n}(t)\right|,
\end{aligned}
\end{equation}
where $\beta$ is the weight factor for leveraging the desired reward and the constraint-based penalty.
\subsubsection{Embedding the Decoupled Softmax Layer into Actor}
\label{sssec:Softmax}
\begin{figure}[t]
	\centering
	\includegraphics[width=0.45\textwidth]{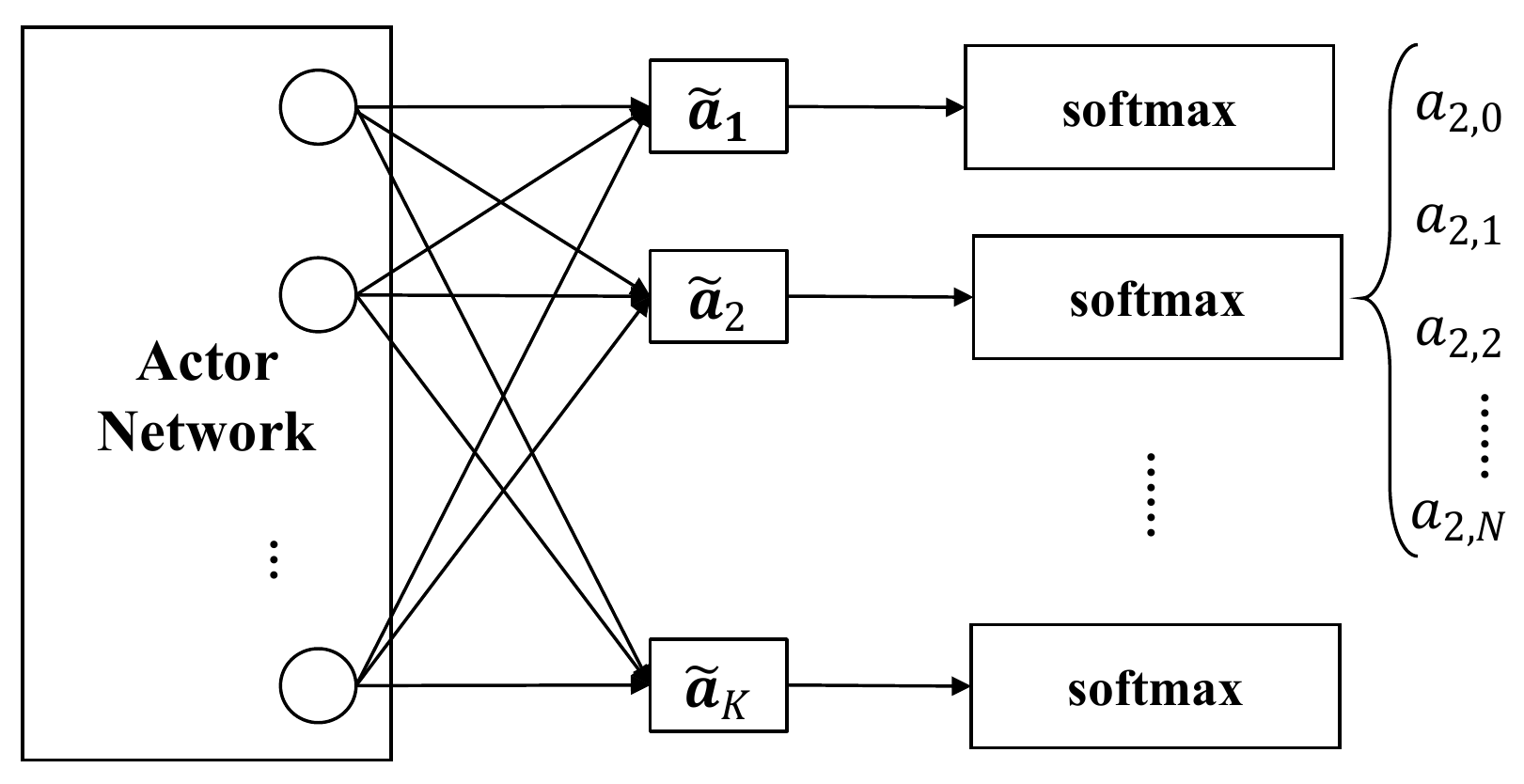}
	\vspace{0ex}
	\caption{Actor's output layer with decoupled softmax activation}
	\label{fig:SepOutAct}
	\vspace{0ex}
\end{figure}
 In this method, we introduce a decoupled regularization layer into the output layer of the actor network, such that this layer becomes part of the end-to-end back propagation training of the neural network. Since the softmax function realizes for each $\va_k$ the following projection
$$
 \sigma:\R^{N+1}\to\left\{\va_k\in\R^{N+1}\Big| a_{k,n}\geq 0, \sum_{n=0}^{N} a_{k,n} = 1\right\},
 $$
 the decoupled softmax layer well addresses the intra-cell inter-slice resource constraints $\sum_{n=0}^{N} a_{k,n} = 1$, $\forall k\in\K$ as shown in Fig. \ref{fig:SepOutAct}.

The benefit of applying the decoupled softmax layer versus the reshaping of reward function is that, because the softmax regularization is part of the end-to-end back propagation, the agent training is usually more stable and converges faster.
\section{Performance Evaluation}\label{sec:numerical}
\begin{figure}[t]
    \centering
    \includegraphics[width=.45\textwidth,height=4.5cm]{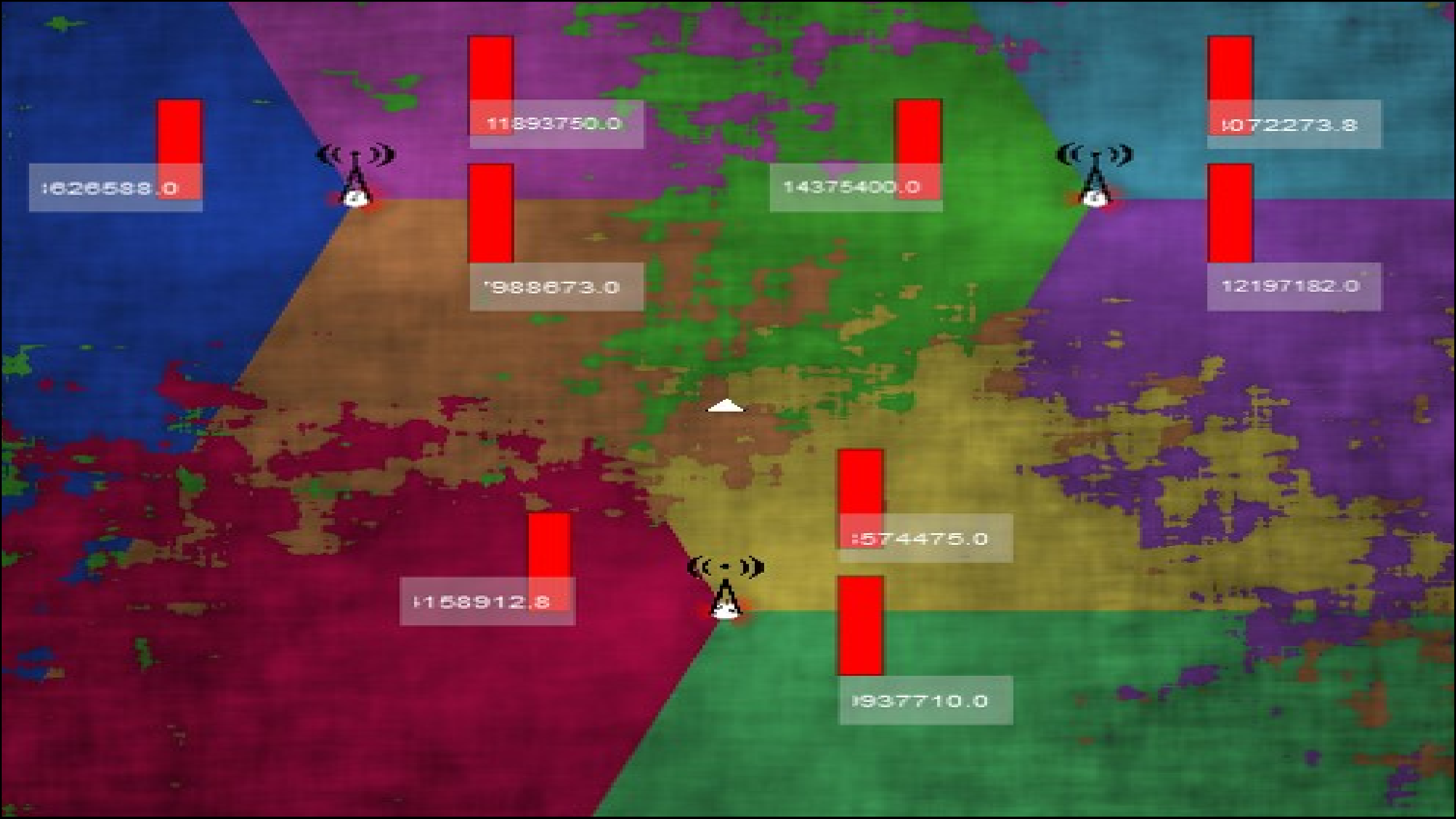}
    \caption{Environment setup for experiments}
    \label{fig:SeaEnv}
\end{figure}

In this section, we evaluate the performance of the proposed distributed schemes for inter-cell slicing resource partitioning introduced in Section \ref{subsec:schemes} with a system-level simulator \cite{SeasonII}, which mimics real-life network scenarios with customized network slicing traffic, user mobility, and network topology. A small urban area of three sites is selected, as demonstrates in Fig. \ref{fig:SeaEnv}. At each three-sector site, three cells are deployed using LTE radio technology with $2.6$ GHz. Thus, we have in total $K=9$ cells. We use the realistic radio propagation model Winner+\cite{Winner}.
The system is built up with $N=2$ network slices: Slice $1$ supporting video traffic and Slice $2$ supporting HTTP traffic. We define slice-specific expected bit rates $\phi_1^* = 5$ MBit/s and $\phi_2^* = 3$ MBit/s respectively and the same network latency requirements $d_n = 1$ ms, $n=1,2$ (due to the current scheduler limitation of the simulator, we can only apply one latency requirement but different throughput requirements). All cells in the network have the same fixed bandwidth $B = 20$ MHz.

\begin{figure}[t]
    \centering
    \vspace{-2ex}
    \includegraphics[width=.45\textwidth]{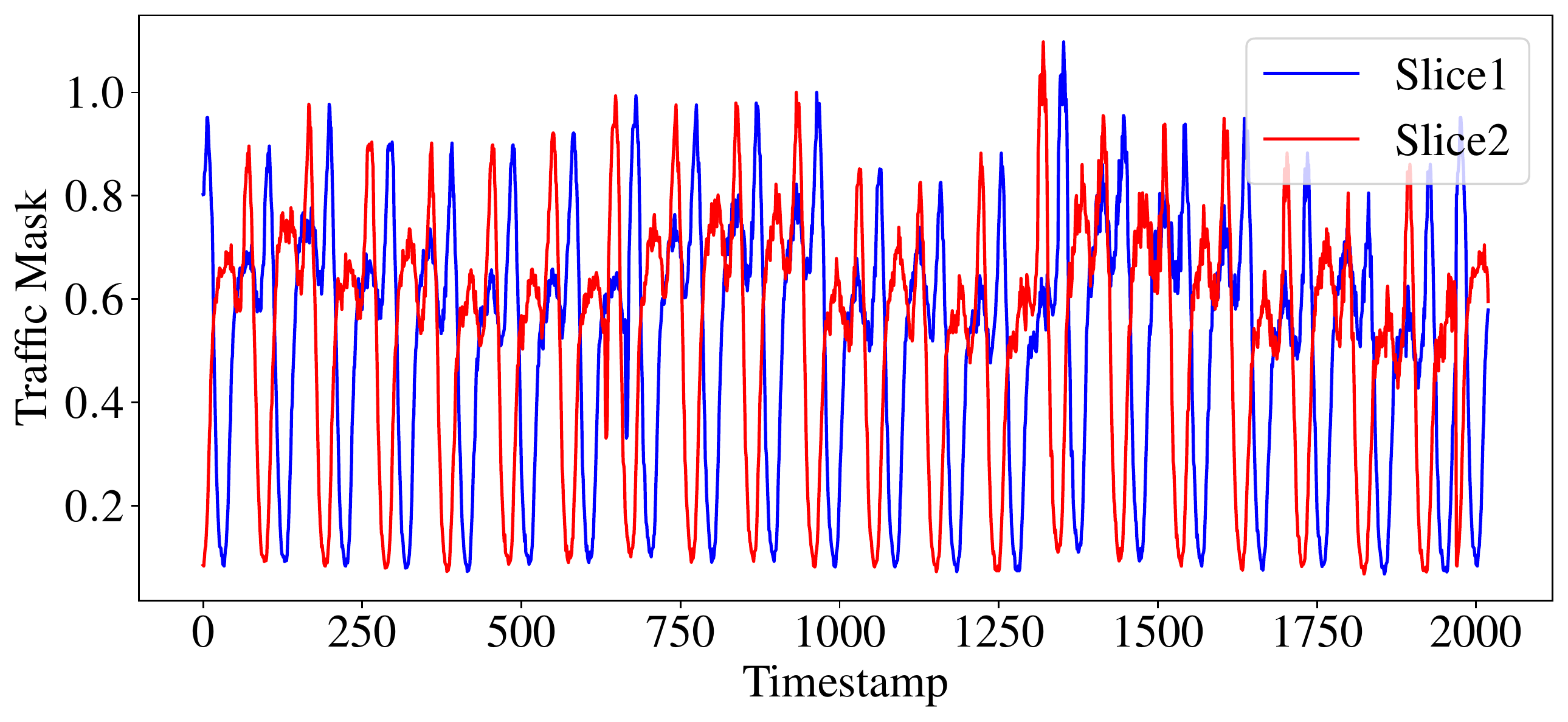}
    \caption{Traffic mask to imitate the time varying network traffic}
    \label{fig:TrafMask}
\end{figure}

We define two groups of \acp{UE} associated to the defined two slices respectively, both with the maximum group size of $32$, and both move uniformly randomly within the playground. To imitate the time-varying traffic pattern, we also apply a time-dependent traffic mask $\tau_n(t)\in[0, 1]$ for each slice $n=1,2$ to scale the total number of \acp{UE} in the scenario, as shown in Fig. \ref{fig:TrafMask}.

\subsubsection{Schemes and Baselines to Compare}
\label{ssec:comparedSchemes}

We compare the proposed distributed \ac{DRL} schemes in Section \ref{sec:Approaches} with the conventional centralized \ac{DRL} approach and a traffic-aware baseline approach. The schemes to evaluate and compare are summarized as follows.
\begin{table*}[ht]
\caption{Comparison of Dimensions of \ac{DRL} Models Used in Simulation}
\label{Table_RL_Compare}
\begin{tabular*}{1.02\textwidth}{@{\extracolsep{\stretch{1}}}*{5}{r}@{}}
\cline{1-4}
\multicolumn{1}{|l|}{}      & \multicolumn{1}{l|}{{\bfseries Centralized}}         & \multicolumn{1}{l|}{{\bfseries Distributed without Coordination}}    & \multicolumn{1}{l|}{{\bfseries Distributed with Coordination}}     &  \\ \cline{1-4}
\multicolumn{1}{|l|}{{\bfseries State}}  & \multicolumn{1}{l|}{Global state $\vs\in\R^{54}$} & \multicolumn{1}{l|}{Local state $\vs_k\in\R^{6}$ }    & \multicolumn{1}{l|}{Local state with extracted message $[\vs_k, \vc_k]\in\R^{8}$} &  \\ \cline{1-4}
\multicolumn{1}{|l|}{{\bfseries Action}} & \multicolumn{1}{l|}{Global action $\va\in [0,1]^{27}$}       & \multicolumn{1}{l|}{Local action $\va_k\in[0,1]^{3}$}   & \multicolumn{1}{l|}{Local action $\va_k\in[0,1]^{3}$}                       &  \\ \cline{1-4}
\multicolumn{1}{|l|}{{\bfseries Reward}} & \multicolumn{1}{l|}{Global reward $r$ in \eqref{reward}}       & \multicolumn{1}{l|}{Local reward $r_k$ in \eqref{reward_dis}} & \multicolumn{1}{l|}{Local reward $r_k$ in \eqref{reward_dis}}                     &  \\ \cline{1-4}
                             &                                          &                                     &                                                         & 
\end{tabular*}
\vspace{-5ex}
\end{table*}

\begin{itemize}
	\item {\bf Cen-Pen}: centralized \ac{DRL} approach with penalized reward as described in Section \ref{sssec:Penalty}. We assume that a single agent has full observation of the global state $\vs\in\Ss$, computes the global reward $r$ based on \eqref{reward}, and makes the decision of the slicing resource partitioning for all agents $\va\in\A$. The dimensions of the centralized and distributed \ac{DRL} models used in the simulation are compared in Table \ref{Table_RL_Compare}.
	\item {\bf Cen-Soft}: same centralized \ac{DRL} approach as Cen-Pen but with embedded softmax layer as introduced in Section \ref{sssec:Softmax}.
	\item {\bf Dist}: distributed multi-agent \ac{DRL} scheme as introduced in Section \ref{sssec:Distributed} with embedded softmax layer.
	\item {\bf Dist-Comm}: coordinated distributed multi-agent \ac{DRL} scheme with inter-cell communication introduced in Section \ref{sssec:Distributed_Comm} and embedded softmax layer.
	\item {\bf Baseline}: a traffic-aware baseline that dynamically adapts to current per-slice traffic amount. In each cell, the resource are split proportionally to the number of active \acp{UE} per slice. 
\end{itemize}

\subsubsection{Hyperparameters used for Learning}
As for \ac{DRL} training, we use \ac{MLP} architecture for actor-critic networks. In Cen-Soft and Cen-Pen schemes, the models of actor-critic networks are both built up with $3$ hidden layers, with the number of neurons $(96, 64, 48)$ and $(120, 64, 32)$, respectively. While for distributed schemes, both actor-critic networks only have two hidden layers as $(48, 24)$ and $(64, 24)$. In all schemes, the learning rate of actor and critic are $0.0005$ and $0.001$ respectively with Adam optimizer and training batch size of $32$. We choose a small \ac{DRL} discount factor $\gamma = 0.1$, since the current action has a strong impact on the instantaneous reward while much less impact on the future. For training setups, we applied $2500$ steps for exploration, $10000$ steps for \ac{DRL} learning and final $2500$ steps for evaluation.

\subsubsection{Performance Comparison}
\begin{figure}[t]
	\centering
	\includegraphics[width=.45\textwidth]{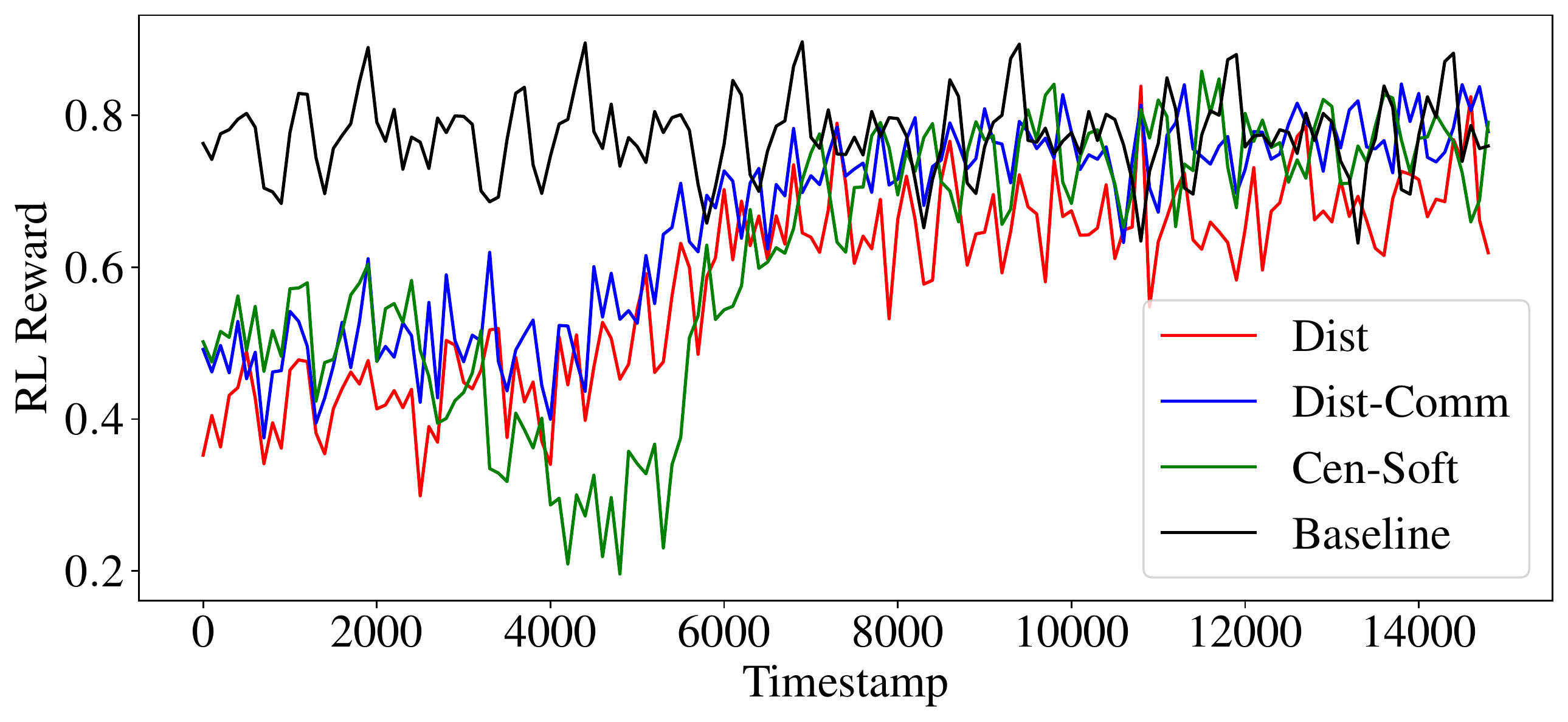}
	\caption{Comparison of reward among schemes}
	\label{fig:Compare_baseline_reward}
\end{figure}

\begin{figure}[t]
	\centering
	\includegraphics[width=.45\textwidth]{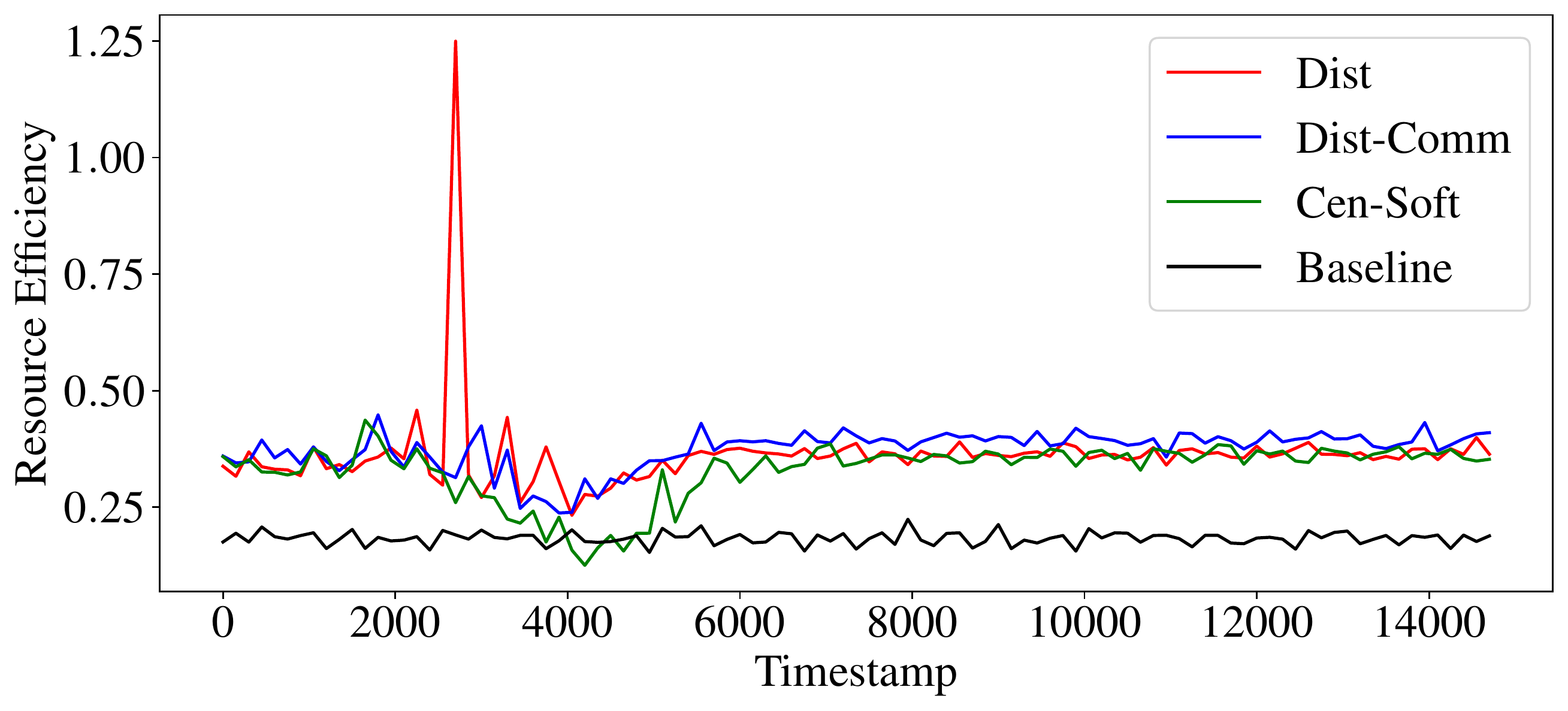}
	\caption{Comparison of resource efficiency among schemes}
	\label{fig:Compare_baseline_effi}
\end{figure}

Fig. \ref{fig:Compare_baseline_reward} demonstrates the comparison of reward defined in (\ref{reward}) during the training process among the schemes Cen-Soft, Dist, Dist-Comm and Baseline defined in Section \ref{ssec:comparedSchemes}, while Fig. \ref{fig:Compare_baseline_effi} compares minimum resource efficiency among slices. The resource efficiency $\eta_k$ for cell $k\in\K$ is given by $\eta_k = (1/N) \sum_{n\in\N} \phi_{k,n}(t)/(a_{k,n}B)$.

As shown in Fig. \ref{fig:Compare_baseline_reward} and \ref{fig:Compare_baseline_effi}, all \ac{DRL} approaches learn to achieve similar service performance to Baseline, while proving more than two-fold increase in resource efficiency by introducing the headroom in action choices. Note that Baseline dynamically captures time-varying traffic pattern and offers all resource to the \acp{UE}, it provides sufficiently good service performance while suffering from low resource efficiency. 

Another observation is that, the proposed Dist-Comm scheme slightly outperforms Cen-Soft within the same training time period. The centralized approach converges slower and often experiences extremely poor performance during training, because it has much higher action and state dimensions and requires longer training to converge to a good solution. In comparison between Dist and Dist-Comm schemes, it is obvious that inter-agent coordination helps Dist-Comm outperform Dist in terms of both service performance and resource efficiency.

\begin{figure}[t]
	\centering
	\vspace{-2ex}
	\includegraphics[width=.45\textwidth]{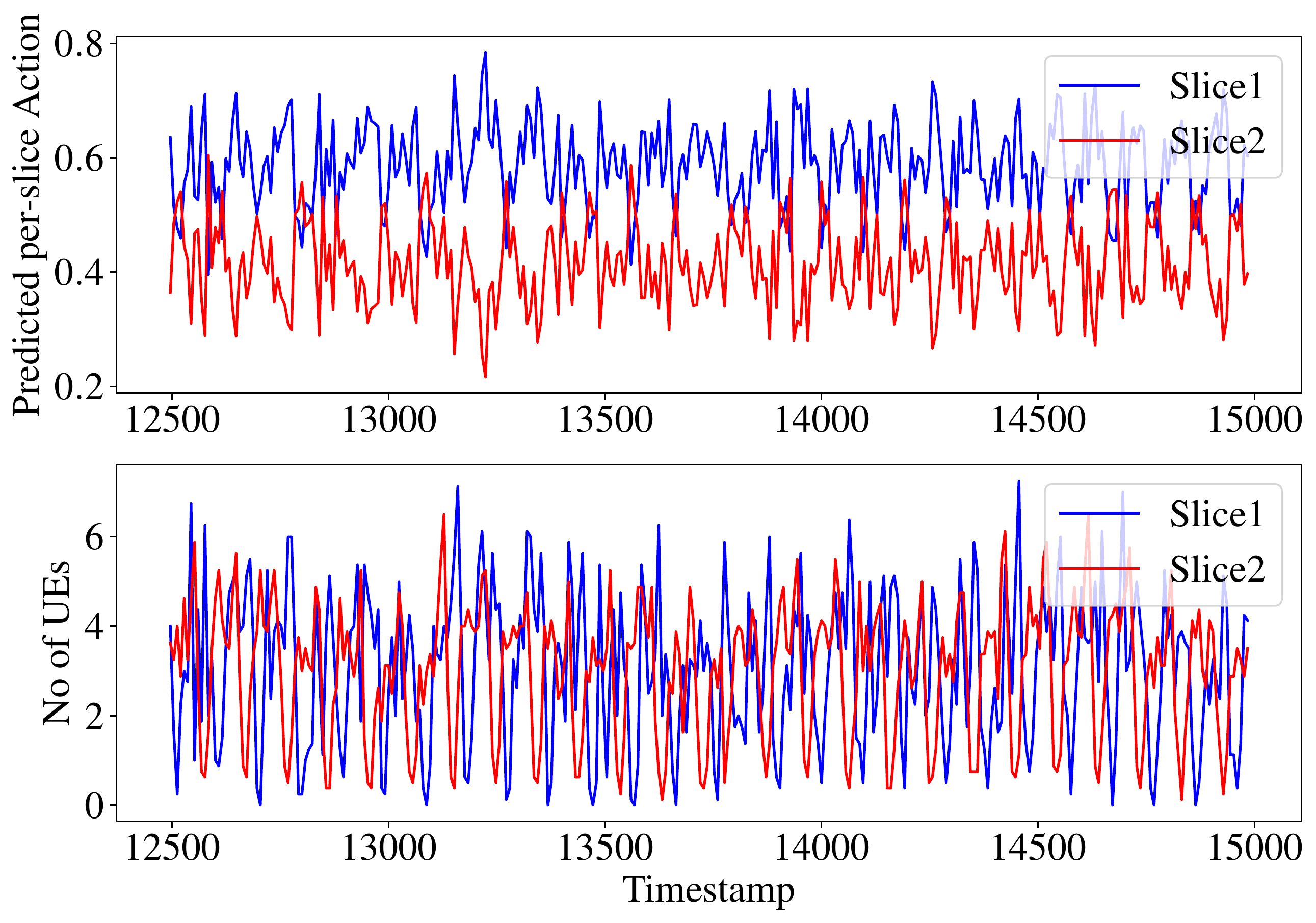}
	\caption{Adaptive action to traffic mask after training}
	\label{fig:Action_UE}
\end{figure}

Fig. \ref{fig:Action_UE} shows the predicted action, i.e., per-slice resource partitioning, and the predefined traffic mask of scheme Cen-Soft in cell $k=9$. It verifies that the \ac{DRL} approach predicts actions that well adapt to network traffic dynamically with respect to different slice-specific throughput requirements.

\begin{figure}[t]
	\centering
	\includegraphics[width=.45\textwidth]{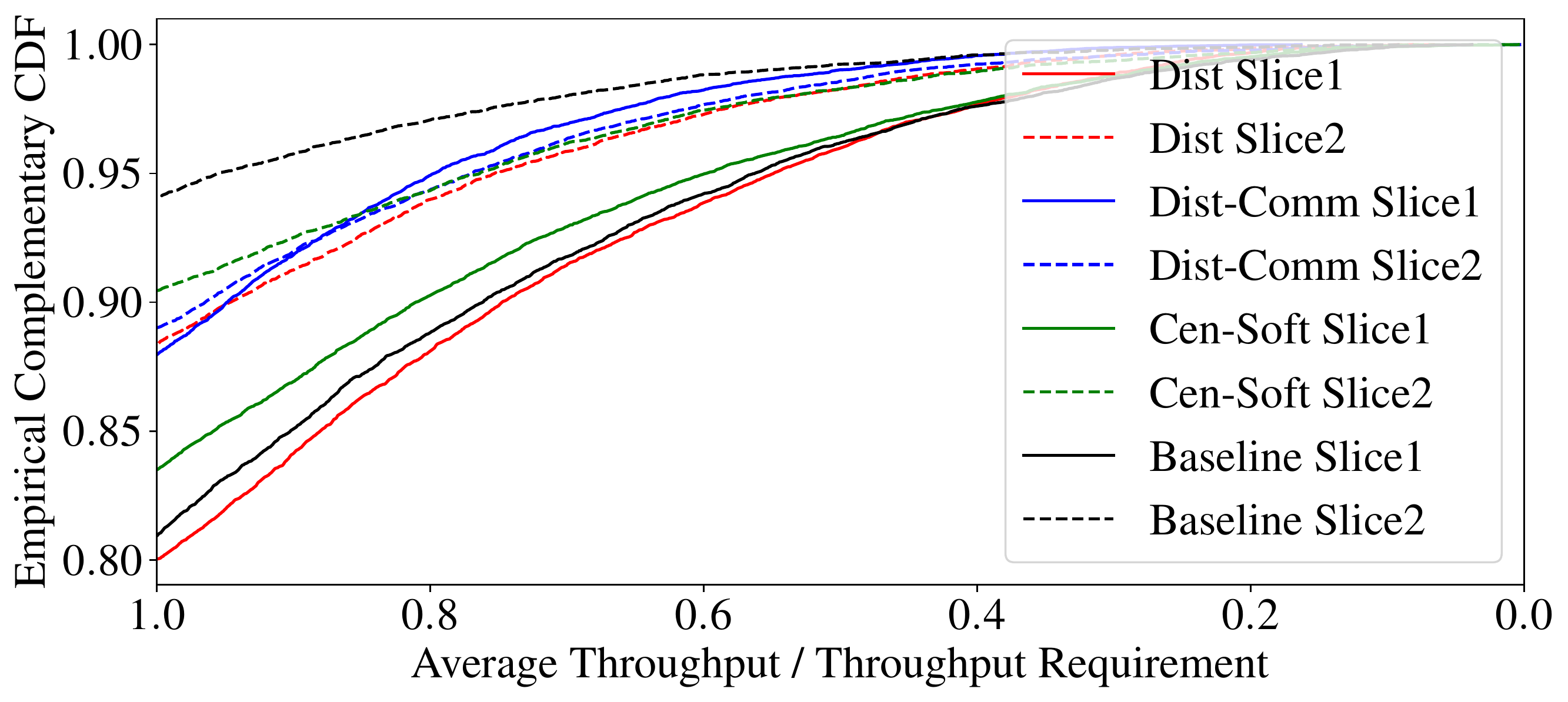}
	\caption{Comparing slice throughput from different approaches}
	\label{fig:Compare_all_thr}
\end{figure}

\begin{figure}[t]
	\centering
	\includegraphics[width=.45\textwidth]{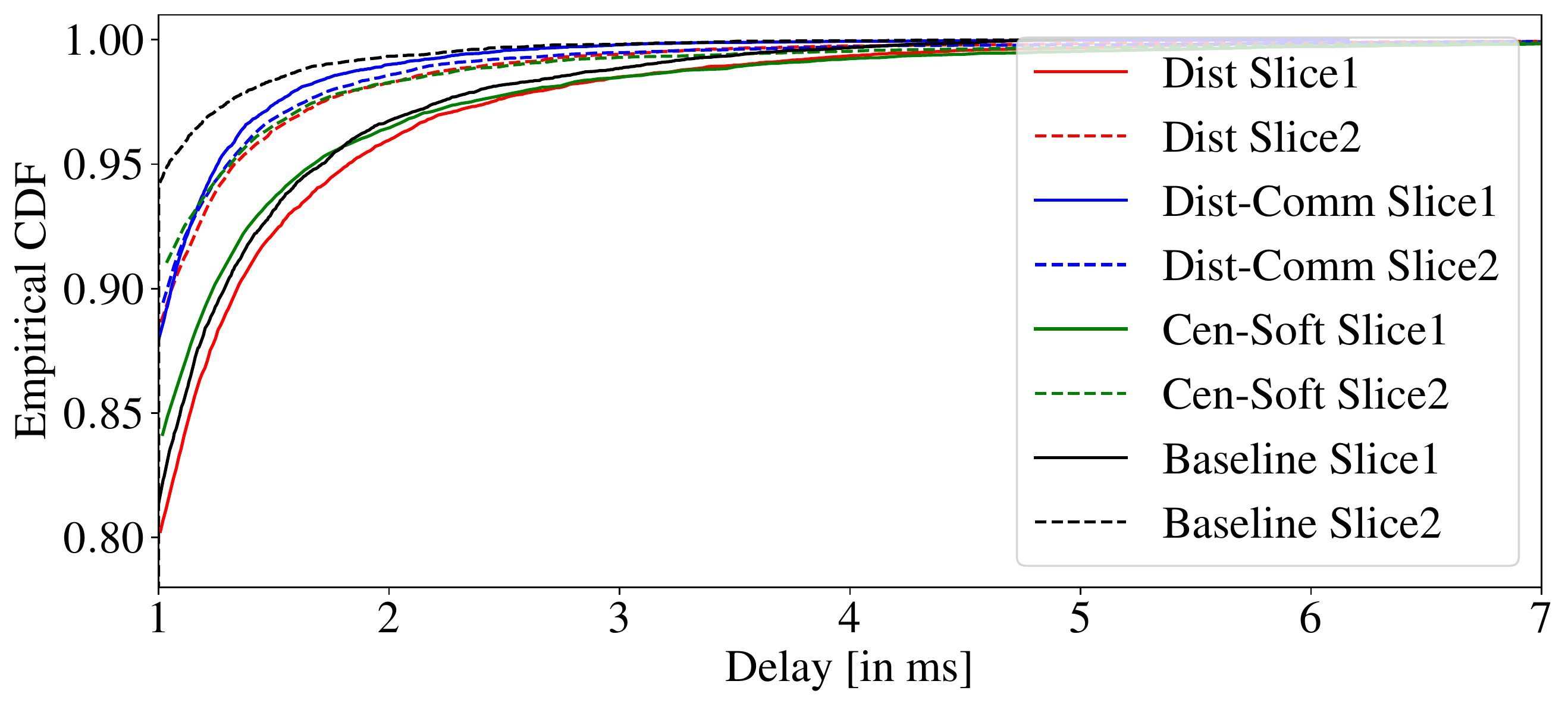}
	\caption{Comparing slice delay from different approaches}
	\label{fig:Compare_all_delay}
\end{figure}

The above-illustrated results show the algorithms' performance in terms of our objectives, i.e, maximizing the minimum service quality among all slices and cells. In the following, let us take a deeper look into the general performance in terms of the service quality distributions. Fig. \ref{fig:Compare_all_thr} illustrates the empirical complementary \ac{CDF} (or called survival function) that equals $1-F_X(x)$ where $F_X(x)$ denotes the \ac{CDF}. We observe that our proposed Dist-Comm achieves best balance between the two slices, with both slices achieving $>88\%$ of the satisfaction ratio with the expected throughput, while Baseline and Cen-Soft provide only $82\%$ and $84\%$ for Slice $1$ respectively. Fig. \ref{fig:Compare_all_delay} illustrates the \ac{CDF} of the slice delay. And similar observation can be made, that the proposed Dist-Comm provides fairly balanced service quality to the two slices.

A summarized comparison of the average performance metrics among all approaches in the testing phase are listed in Table \ref{Table_Value_Compare}. We can see that Dist-Comm provides the best performance in terms of the desired reward, resource efficiency, and the throughput and delay requirements. Moreover, it encourages a more balanced service quality between the two slices. 

\begin{table*}[ht]
\centering
\caption{Compare of Average Performance Metrics among Different Approaches}
\label{Table_Value_Compare}
\begin{tabular*}{1.05\textwidth}{@{\extracolsep{\stretch{1}}}*{6}{r}@{}}
\cline{1-5}
\multicolumn{1}{|l|}{}  & \multicolumn{1}{l|}{{\bf Dist}}   & \multicolumn{1}{l|}{{\bf Dist-Comm}}    & \multicolumn{1}{l|}{{\bf Cen-Soft}}  & \multicolumn{1}{l|}{{\bf Baseline}}     &  \\ 
\cline{1-5}
\multicolumn{1}{|l|}{{\bf RL Reward}}  & \multicolumn{1}{l|}{0.697} & \multicolumn{1}{l|}{{\bf 0.775}}    & \multicolumn{1}{l|}{0.756} & \multicolumn{1}{l|}{0.771} &  \\ 
\cline{1-5}
\multicolumn{1}{|l|}{{\bf Resource Efficiency}} & \multicolumn{1}{l|}{0.367}       & \multicolumn{1}{l|}{{\bf 0.374}}   & \multicolumn{1}{l|}{0.362}   & \multicolumn{1}{l|}{0.183}   &  \\ 
\cline{1-5}
\multicolumn{1}{|l|}{{\bf Per-Slice Throughput / Requirement}} & \multicolumn{1}{l|}{(0.940, 0.969)}       & \multicolumn{1}{l|}{({\bf 0.975}, 0.972)} & \multicolumn{1}{l|}{(0.950, 0.972)}     & \multicolumn{1}{l|}{(0.942, {\bf 0.985})}                &  \\ 
\cline{1-5}
\multicolumn{1}{|l|}{{\bf Per-Slice Delay (ms)}}  &   \multicolumn{1}{l|}{(1.14, 1.06)}                 &  \multicolumn{1}{l|}{({\bf 1.04}, 1.11)} &    \multicolumn{1}{l|}{(1.15, 1.07)}                           & \multicolumn{1}{l|}{(1.15, {\bf 1.03})}  &\\ 
\cline{1-5}
                     &                 &              &                                                         & 
\end{tabular*}
\vspace{-5ex}
\end{table*}

\begin{figure}[t]
	\centering
	\includegraphics[width=.45\textwidth]{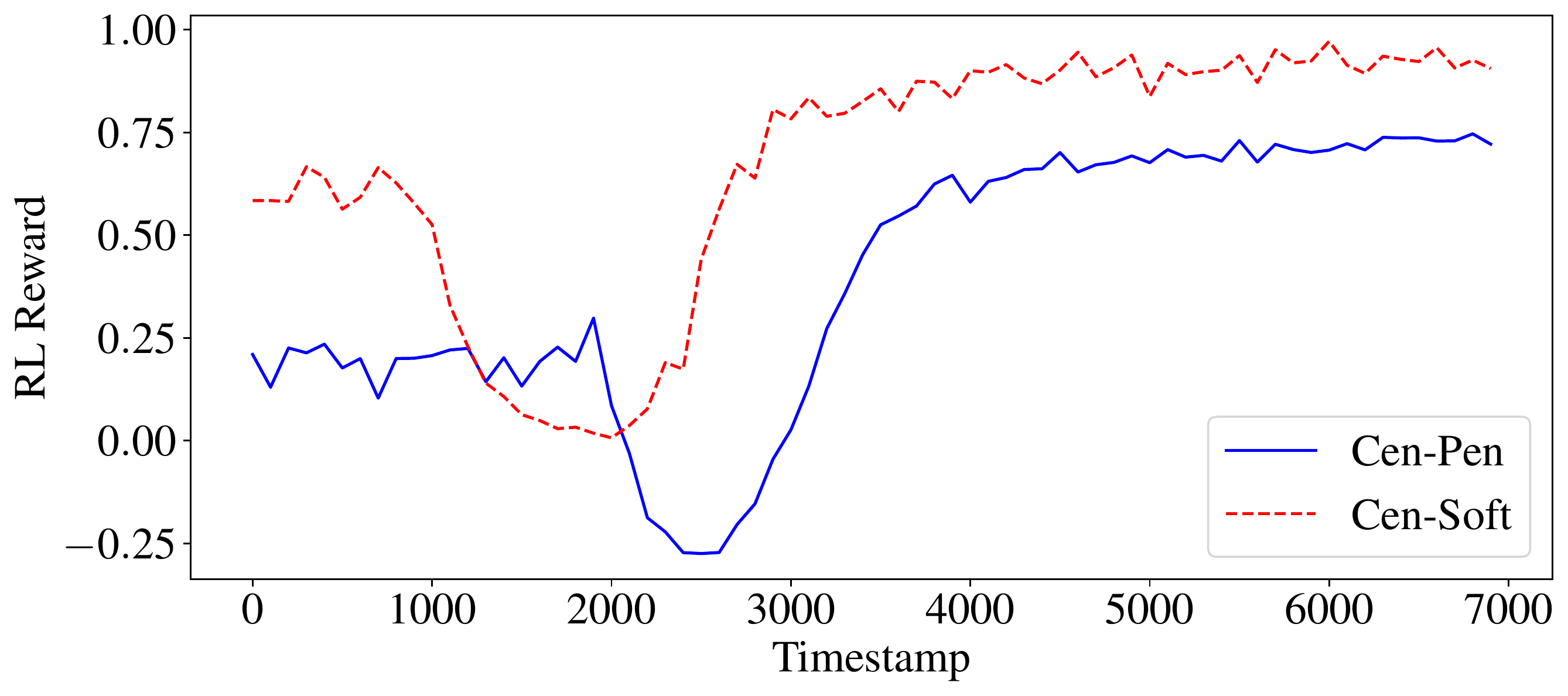}
	\caption{Comparing solutions to \ac{DRL} with resource constraints}
	\label{fig:Reward_Compare}
\end{figure}

%
Last but not least, Fig. \ref{fig:Reward_Compare} illustrates the comparison between the solutions to resource constraints. The embedded softmax layer demonstrates a better performance than the reward shaping. It is also worth mentioning that the results shown in Fig. \ref{fig:Reward_Compare} was obtained with a different smaller environment consisting of $6$ cells, with first $1000$ timestamps for exploration, $6000$  for training and final $1000$ for testing, while with $9$ cells we have difficulties to obtain converging results using reward shaping. Thus, a hypothesis is that the shaped reward function is more complex, and easily causes oscillating and unstable training experience. 

\subsubsection{Key Takeaways}
In the following we summarize the takeaways from our numerical analysis:
\begin{itemize}
    \item Both centralized and distributed \ac{DRL}-based approaches demonstrate good learning capability for adapting to slice-aware traffic and providing good service quality. Moreover, due to the introduction of the headroom, they provide more than two-fold increase in resource efficiency compared to the traffic-aware baseline.
    \item The distributed coordinated scheme achieves better performance than the centralized approach when both are trained with same limited time period. Introducing inter-agent coordination and letting the multiple agents share load information help improve the performance of the distributed scheme, while remaining lower model complexity and faster convergence compared to the centralized approach. A further benefit is that it achieves a more balanced service quality among different slices.
    \item When dealing with inter-slice resource constraints, embedding decoupled softmax layer outperforms reward shaping in terms of faster convergence and preventing deep oscillating during training.
\end{itemize}
\section{Conclusion}\label{sec:concl}
In this paper, we formulated the dynamic inter-cell slicing resource partitioning problem to meet the slice-aware service requirements and improve the resource efficiency by jointly optimizing the inter-cell inter-slice resource partitioning and resource headroom. We proposed a distributed multi-agent \ac{DRL} solution to solve the problem and compare two different schemes with and without inter-agent coordination. We also proposed two methods, i.e., reward shaping and decoupled softmax embedding,  to allow the \ac{DRL} agents aware of the inter-slice resource constraints. We evaluated the proposed solutions extensively with a system-level simulator and show that the coordinated distributed scheme provides better slice-aware service performance than the centralized approach with the same limited training time, while achieving more than two-fold increase in resource efficiency compared to the traffic-aware baseline.

\bibliographystyle{IEEEtran}
\bibliography{myreferences}

\end{document}

%% file: macro.tex
\usepackage{acronym}
\input{acronyms.tex}
\usepackage[left=1.57cm,right=1.57cm,top=0.95cm,bottom=2.54cm]{geometry}
\usepackage[font=small,skip=2pt]{caption}
\usepackage{cite}
\usepackage{amsmath,amssymb,amsfonts}
\usepackage[mathscr]{euscript}
\usepackage{algorithmic}
\usepackage{graphicx}
\usepackage{textcomp}
\usepackage{xcolor}
\def\BibTeX{{\rm B\kern-.05em{\sc i\kern-.025em b}\kern-.08em
    T\kern-.1667em\lower.7ex\hbox{E}\kern-.125emX}}

\usepackage{url}
\usepackage{bbm}

\newtheorem{problem}{Problem}

\DeclareMathOperator*\argmax{arg \, max}		

\newcommand{\field}[1]{\mathbb{#1}}

\newcommand{\set}[1]{\mathcal{#1}}

\newcommand{\R}{{\field{R}}}  
\newcommand{\NN}{{\field{N}}}
\newcommand{\Ex}{{\field{E}}}

\newcommand{\ve}[1]{\boldsymbol{\mathbf{#1}}} 

\newcommand{\vs}{\ve{s}}

\newcommand{\vm}{\ve{m}}

\newcommand{\vc}{\ve{c}}

\newcommand{\va}{\ve{a}}

\newcommand{\N}{{\set{N}}}

\newcommand{\Ss}{{\set{S}}}

\newcommand{\K}{{\set{K}}}
\newcommand{\A}{{\set{A}}}

%% file: acronyms.tex
\acrodef{3GPP}{3rd generation partnership project}
\acrodef{3D}{three-dimensional}
\acrodef{2D}{two-dimensional}
\acrodef{5G}{fifth generation}
 \acrodef{ABS}{almost blank subframe}
 \acrodef{AES}{advanced encryption standard}
\acrodef{APE}{absolute percentage error}
\acrodef{AR}{augmented reality}
    \acrodef{BS}{base station}
		\acrodef{BA}{bundle adjustment}
    \acrodef{CDF}{cumulative distribution function}
		\acrodef{CEVP}{constrained eigenvalue problem}
    \acrodef{CSI}{channel state information}
    \acrodef{CQI}{channel quality indicator}
		\acrodef{CNN}{convolutional neural network}
	\acrodef{CMDP}{Constrained Markov Decision Process}
	\acrodef{CDF}{cumulative distribution function}
\acrodef{DL}{downlink}
    \acrodef{DUDe}{downlink and uplink decoupling}
		\acrodef{DoF}{degree of freedom}
		\acrodef{DNN}{deep neural network}
		\acrodef{DDPG}{Deep Deterministic Policy Gradient}
\acrodef{DRL}{deep reinforcement learning}
\acrodef{eICIC}{enhanced intercell interference coordination}
\acrodef{ESD}{energy spectral density}
\acrodef{ECDSA}{elliptic curve digital signature algorithm}
\acrodef{FDD}{frequency division duplex}
    \acrodef{FDMA}{frequency division multiple access}
		\acrodef{fps}{frame per second}
   \acrodef{GP}{Gaussian process}
    \acrodef{GPS}{global positioning system}
		\acrodef{GUI}{graphical user interface}
\acrodef{HetNet}{heterogeneous network}
\acrodef{HOG}{histogram of oriented gradients}
    \acrodef{ICI}{inter-cell interference}
		\acrodef{IMI}{inter-mode interference}
		\acrodef{IMU}{inertial measurement unit}
		\acrodef{IoT}{internet of things}
\acrodef{KL}{Kullback-Leibler}
\acrodef{KPI}{Key Performance Indicator}
\acrodef{LTE}{long term evolution}

\acrodef{mAP}{mean average precision}
\acrodef{MARL}{multi-agent reinforcement learning}
\acrodef{MAC}{media access control}
\acrodef{MAPE}{mean absolute percentage error}
\acrodef{MIMO}{multiple-input and multiple-output}
\acrodef{MRU}{minimum resource unit}
\acrodef{mmWave}{millimeter wave}
\acrodef{MLP}{multi-layer perception}
\acrodef{MILP}{mixed integer linear programming}
\acrodef{MDP}{Markov Decision Process}
\acrodef{MMDPs}{Multi-Agent Markov Decision Process}

\acrodef{NLES}{nonlinear equation system}
   \acrodef{OFDM}{orthogonal frequency division multiplexing}
	\acrodef{ORB}{oriented FAST and rotated BRIEF}
	\acrodef{OAM}[O$\&$M]{network operation and maintenance}
    \acrodef{PDF}{probability density function}
    \acrodef{PHY}{physical layer}
		\acrodef{PSD}{power spectral density}
    \acrodef{PRB}{physical resource block}
	\acrodef{PPO}{Proximal Policy Optimization}
   \acrodef{QoE}{quality of experience}
    \acrodef{QoS}{quality of service}
    \acrodef{RAN}{radio access network}
		\acrodef{RANSAC}{random sample consensus}
		\acrodef{RB}{resource block}
		\acrodef{RBS}{removal of bottleneck services}
		\acrodef{R-FCN}{region-based fully convolutional networks}
		\acrodef{RL}{Reinforcement Learning}
		\acrodef{RMDI}{resource muting for dominant interferer}
		\acrodef{RMSD}{root mean square distance}
		\acrodef{ROI}{region of interest}
		\acrodef{RPN}{region proposal network}
    \acrodef{RRM}{radio resource management}
		\acrodef{RU}{resource unit}
		\acrodef{RX}{receiver}
 \acrodef{SAFP}{successive approximation of fixed point}
 \acrodef{SE}{secure element}
    \acrodef{SDN}{software defined network}
    \acrodef{SNR}{signal-to-noise ratio}
    \acrodef{SINR}{signal-to-interference-plus-noise ratio}
\acrodef{SIR}{signal-to-interference ratio}
\acrodef{SIF}{standard interference function}
\acrodef{SLAM}{simultaneous localization and mapping}
\acrodef{SLAMORE}{Simultaneous Localization and Mapping with Object REcognition}
\acrodef{SoC}{system-on-chip}
\acrodef{SON}{self-organizing networks}
\acrodef{SSD}{single-shot multibox detector}
\acrodef{SPI}{serial peripheral interface}
    \acrodef{SVM}{support vector machine}
		\acrodef{SVD}{singular value decomposition}
    \acrodef{TCP}{transmission control protocol}
		\acrodef{TLS}{transport layer security}
		\acrodef{TDD}{time division duplex}
    \acrodef{TDMA}{time division multiple access}
		\acrodef{TTI}{transmission time interval}
		\acrodef{TX}{transmitter}
    \acrodef{TDL}{temporal different learning}
	\acrodef{TD3}{Twin Delayed Deep Deterministic policy gradient algorithm}
	\acrodef{TRPO}{Trust Region Policy Optimization}
\acrodef{UART}{universal asynchronous receiver-transmitter}
\acrodef{UDP}{user datagram protocol}
\acrodef{UE}{user equipment}
\acrodef{UI}{user interface}
\acrodef{UL}{uplink}
\acrodef{UAV}{unmanned aerial vehicle}
\acrodef{VO}{visual odometry}
    \acrodef{WLAN}{wireless local area network}
\acrodef{YOLO}{you only look once}